\documentclass[reprint,superscriptaddress,amsmath,amssymb,prl]{revtex4-1}

\usepackage{graphicx}
\usepackage{subfigure}
\usepackage{dcolumn}
\usepackage{bm}
\usepackage{xcolor}%
\usepackage{amsmath}
\usepackage{verbatim}
\usepackage{mathtools}
\usepackage{epstopdf}

\def\nin{\noindent} 
\def\beq{\begin{equation}}
\def\eeq{\end{equation}}

\begin{document}

\title{Structural Evolution of Granular Systems: Theory}

   \author{Clara C. Wanjura}\email{ccw45@cam.ac.uk}
   \affiliation{Ulm University, Albert-Einstein-Allee 11, 89081 Ulm, Germany}
   \affiliation{Cavendish Laboratory, Cambridge University, \\
                JJ Thomson Avenue, Cambridge CB3 0HE, UK}
    \author{Paula Gago}
   \affiliation{Imperial College London, London SW7 2AZ, UK}
   \author{Takashi Matsushima}
   \affiliation{University of Tsukuba, Japan}
 \author{Raphael Blumenfeld}\email{rbb11@cam.ac.uk}
    \affiliation{Cavendish Laboratory, Cambridge University, \\
                JJ Thomson Avenue, Cambridge CB3 0HE, UK}
   \affiliation{Imperial College London, London SW7 2AZ, UK}


\begin{abstract}
A general theory is developed for the evolution of the cell order (CO) distribution in planar granular systems. Dynamic equations are constructed and solved in closed form for several examples: systems under compression; dilation of very dense systems; and the general approach to steady state. We find that all the steady states are stable and that they satisfy detailed balance-like condition when the CO$\,\leq 6$. Illustrative numerical solutions of the evolution are shown. Our theoretical results are validated against an extensive simulation of a sheared system. The formalism can be readily extended to other structural characteristics, paving the way to a general theory of structural organisation of granular systems.

\end{abstract}


\maketitle

\nin {\bf Introduction:}
Understanding and modelling self-organisation of dense granular matter (DGM) under external forces is essential to many natural phenomena and technological applications. Examples are: consolidation and failure of soils, packing of particulates in technological processes, initiation of avalanches, flow of slurries, and dense colloidal suspensions, to mention a few. 
This is also one of the most important problems in granular science \cite{EdOa89a, Vogel_Roth_2003, Cheng_etal_1999, BaBl02, Aste_etal_2007, Meyer-etal2010} because both the dense flow dynamics and the large-scale properties of consolidated DGM depend strongly on the particle-scale structure. Such properties are: permeability to fluid flow \cite{Vogel_Roth_2003}, catalysis, heat exchange, and functionality of fuel cell electrodes \cite{Ametal17}.  
In this paper we address this generic problem and develop a set of equations for the quasi-static structural evolution of two-dimensional (2D) systems of rigid grains.

Structural evolution of DGM is mediated by continual making and breaking of intergranular contacts. In turn, the temporal structural configuration determines the force transmission in the medium, which drives the evolution. The intergranular contact network can be regarded as a graph containing voids (or cells), each of which characterized by the number of grains enclosing it -- its order. Developing a theory for the evolution of the cell order distribution (COD) is the specific aim here. This distribution has been argued \cite{Fretal08} and shown \cite{MaBl14,MaBl17} to converge to a universal form, in which the intergranular friction is scaled away. 
Specifically, a set of evolution equations is constructed and solved in closed form, under some assumptions, both for very dense closed systems and for closed systems approaching a limit steady state. More general cases are solved numerically. We then compare the theoretical results with numerical simulations of sheared systems and show that there is good agreement.
\begin{figure}
\includegraphics[width=.45\textwidth]{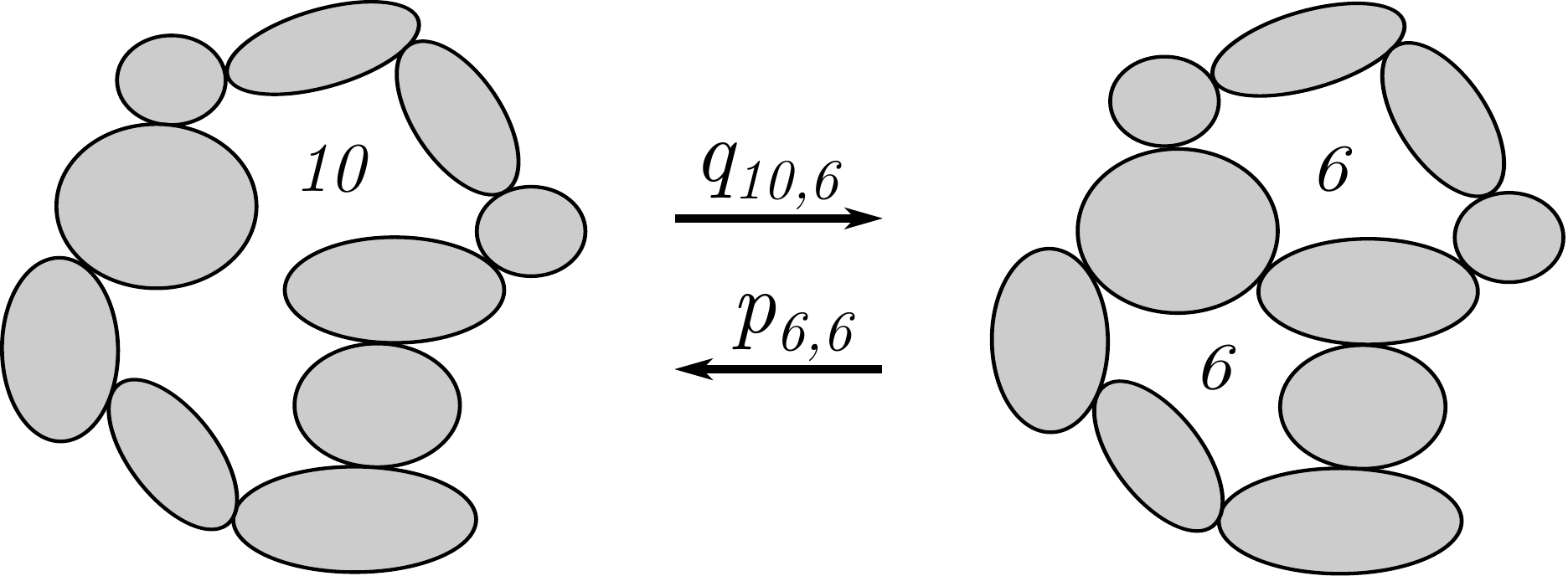}%
\caption{\label{Fig1} 
Making a contact, which occurs at rate $q_{10,6}$, splits a $10$-cell into two $6$-cells and vice versa at rate $p_{6,6}$.}
\end{figure}%

\nin {\bf The evolution equations:}
When a 2D DGM undergoes slow quasi-static dynamics, it is possible to identify, at every moment, a contact network. In this graph the grain centres (however defined) are the vertices and edges connect grains in contact, with each edge corresponding to a specific contact. These make cells and we assign each a cell order defined as the number of grains surrounding it. In such dynamics, the stress response is faster than any other process and a static stress state is established practically immediately after a contact is made or broken. The systems can be regarded then as moving from one stress state to another.

We consider systems free of gravity and body forces. In the concluding discussion, we argue that including these simplifies the following formalism.
The structure evolves through making and breaking of these contacts, which we call contact events (CEs). The former splits a cell into two smaller ones and the latter merges two cells into a larger one. 
If a CE involves two rattler-free neighbour cells of orders $i$ and $j$, the two processes satisfy $i+j\xrightleftharpoons{ } i+j-2$, exemplified in Fig.~\ref{Fig1}. 
Analogously, if the mother cell contains a rattler that participates in the event, the process sum rule is $i+j\xrightleftharpoons{ } i+j-4$.
To model the dynamics of the COD, we define $n_k$ as the total number of $k$-cells in the system and their fraction of the total cell population, $N_c$, as $Q_k = n_k/N_c$. 
We also define the following rates, which we assume are system size-independent: 
$p_{i,j}$ = the merging rate of an $i$- and a $j$-cells into an $i+j-2$-cell; 
$q_{k,i}$ = the  splitting rate of a $k=i+j-2$-cell into an $i$- and a $k-i+2$-cells; 
$r_{i,j}$ = the merging rate of an $i$- and a $j$-cells, containing a rattler, into a $k=i+j-4$-cell; and 
$s_{k,i}$ = the  splitting rate of a $k$-cell, containing a rattler, into an $i$- and a $(k-i+4)$-cell.
The COD evolves via four basic CEs. Two of creation: a $k$-cell is either a merger of $j$ and $k-j+2$, or an offspring of a split large-order cell, and two of annihilation: either by splitting into two offsprings or merging to make a larger cell. 
Each of these processes has an equivalent when the combined cell contains a rattler, in which case the rates $p_{j,k-j+2}$ and $q_{k,j}$ are replaced, respectively, by the rates $r_{j,k-j+4}$ and $s_{k,j}$. 

If very large cells are rare then the occurrence of cells containing more than one rattler is very low and we ignore such events in the following. Then the evolution equations are:
\begin{eqnarray}
\dot{n_k} & = &\frac{1}{2} \sum_{i=3}^{k-1}\big\{ \left[ n_in_{k-i+2}\tilde{p}_{i,k-i+2} - n_k \tilde{q}_{k,i}\right]\left(1 + \delta_{i,k-i+2}\right) + \nonumber \\
& + & \left[n_in_{k-i+4}\tilde{r}_{i,k-i+4} - n_k \tilde{s}_{k,i}\right]\left(1 + \delta_{i,k-i+4}\right)\big\} + \nonumber \\
& + & \sum_{i=k+1}^{\infty} \big\{\left[n_i \tilde{q}_{i,k} - n_kn_{i-k+2}\tilde{p}_{k,i-k+2}\right]\left(1+ \delta_{i,2k-2}\right)  + \nonumber \\
& + & \left[n_i \tilde{s}_{i,k} - n_kn_{i-k+4}\tilde{r}_{k,i-k+4}\right]\left(1+\delta_{i,2k-4}\right) \big\} \text{.}
\label{eqmaster1}
\end{eqnarray}
The 1/2 factor corrects for double counting and the terms containing $\delta$-functions account for loss (gain) of two same order cells upon creation (annihilation) of a larger cell.  The 'tilded' parameters in (\ref{eqmaster1}) are related directly to the size-independent rates: $\tilde p_{i,j}\equiv p_{i,j}/N_c$, $\tilde q_{k,i}\equiv q_{k,i}/N_c$, $\tilde r_{i,j}\equiv r_{i,j}/N_c$, $\tilde s_{k,i}\equiv s_{k,i}/N_c$.
To simplify the following analysis, we ignore rattlers and set $r_{i,j}=s_{i,j}=0$. Including the more realistic rattler-related terms is straightforward, but it would result in cumbersome expressions without adding any more insight. This amounts to assuming that the average number of rattlers per cell type is constant and therefore so is the number of non-rattlers. This assumption is indeed borne out by our numerical simulation results. \\
From eqs. (\ref{eqmaster1}), we note that $\sum_k (k-2) n_k = E$ is a conserved quantity. This quantity has a physical interpretation: $\sum_k n_k \equiv N_c$ and $\sum_k k n_k\equiv Nz$ is twice the number of contacts, with $z$ as the mean coordination number and $N$ as the number of grains. We use Euler's topological expression in the plane to relate the numbers of vertices (=$N$ grains), edges (=$N z/2$), and cells, $N_c$, $N-Nz/2+N_c = \mathcal{O}(\sqrt{N})$,
in which $\mathcal{O}(\sqrt{N})$ are boundary terms. It follows that $E=2N+\mathcal{O}(\sqrt{N})$. Embedding the system on the surface of a sphere, $E=2(N-1)$ exactly.

To make (\ref{eqmaster1}) size-independent, we convert them for the fractions $Q_k$. To this end, we define the deviation of the rattler-free steady state process $i+j\xrightleftharpoons{ } i+j-2$ from `detailed balance'-like steady state,
\beq
\eta_{i,j} \equiv p_{i,j} Q_i Q_j - q_{i+j-2,i} Q_{i+j-2}  \ .
\label{eqEta}
\eeq
When $\eta_{i,j}=0$, the `reaction' $i+j \to i+j-2$ and `back-reaction' $i+j-2 \to i+j$ occur at the same rate, satisfying the definitions of Lewis and Ter Haar for detailed balance in thermodynamic systems \cite{DBL}. The sign of $\eta_{i,j}$ determines which direction of the process $i+j\leftrightharpoons i+j-2$ is more frequent.  Generically, dilation or compression correspond to positive and negative $\eta_{i,j}$, respectively.
Similarly, $\zeta_{i,j} \equiv r_{i,j} Q_i Q_j - s_{i+j-4,i} Q_{i+j-4}$ is the deviation of the rattler-involving steady state process $i+j\xrightleftharpoons{ } i+j-4$ from a detailed balance-like state.

Noting that $\dot{Q_k}=\left(\dot{n_k}-Q_k\dot{N_c}\right)$, the equations for the cell fractions become
\begin{alignat}{2}
   \dot{Q_k} & = &&\frac{1}{2}\sum_{i=3}^{k-1} \eta_{i,k-i+2} \left(1 + \delta_{i,k-i+2}\right) - \nonumber \\
   & && - \sum_{i=k+1}^{\infty} \eta_{k,i-k+2} \left(1+ \delta_{i,2k-2}\right) + Q_k \hspace{-1em}\sum_{\text{all possible}\atop\text{processes $i,j$}}\hspace{-1em}\eta_{i,j} \ ,
\label{eqmaster2}
\end{alignat}
in which we used
\begin{equation}
  \dot N_c = \sum_{k=3}^{\infty} \dot n_k = -N_c\hspace{-1em}\sum_{\text{all possible}\atop\text{processes $i,j$}} \hspace{-1em}\eta_{i,j} \ .
\label{eqNc}
\end{equation}
Eqs. (\ref{eqmaster2}) are now conveniently system size-independent.
In practice, cell orders in realistic quasi-static systems cannot exceed an upper bound $\mathcal{C}$ and remain mechanically stable.

Focusing on closed systems with $N(\gg 1)$ constant, we use Euler's relation again to derive a relation between the rates of change of $N_c$ and $z$: 
\begin{equation}
 \frac{\dot N_c}{N_c} = \frac{\dot{z}}{z-2} \ .
\label{eqNcz}
\end{equation}

\begin{figure}[htbp]
   \centering
   \includegraphics[width=.45\textwidth]{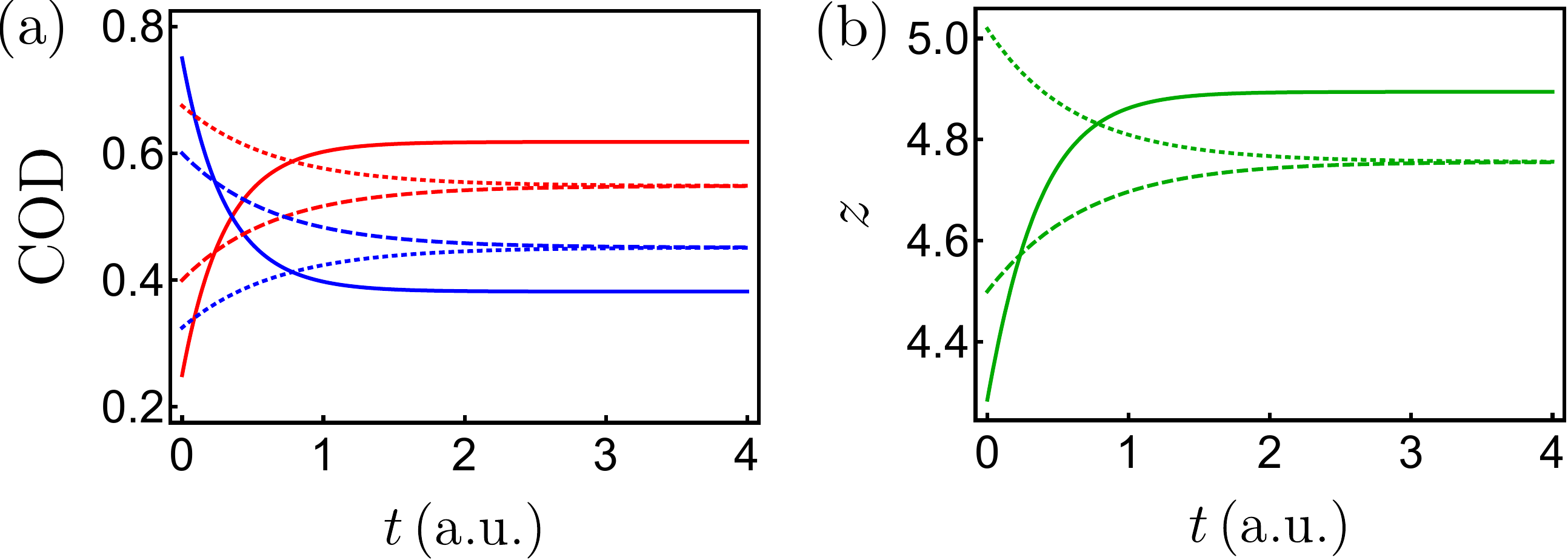}
   \caption{Evolution of the COD of the $3-4$ system for rates: $p_{3,3}=0.6$ and $q_{4,3}=0.4$. (a) $Q_3$ (red) and $Q_4$ (blue); (b) $z$. Starting from three different initial states, the solutions converge to the same steady state.}
   \label{fig:example34Ini}
\end{figure}

\begin{figure}[htbp]
      \centering
      \includegraphics[width=.45\textwidth]{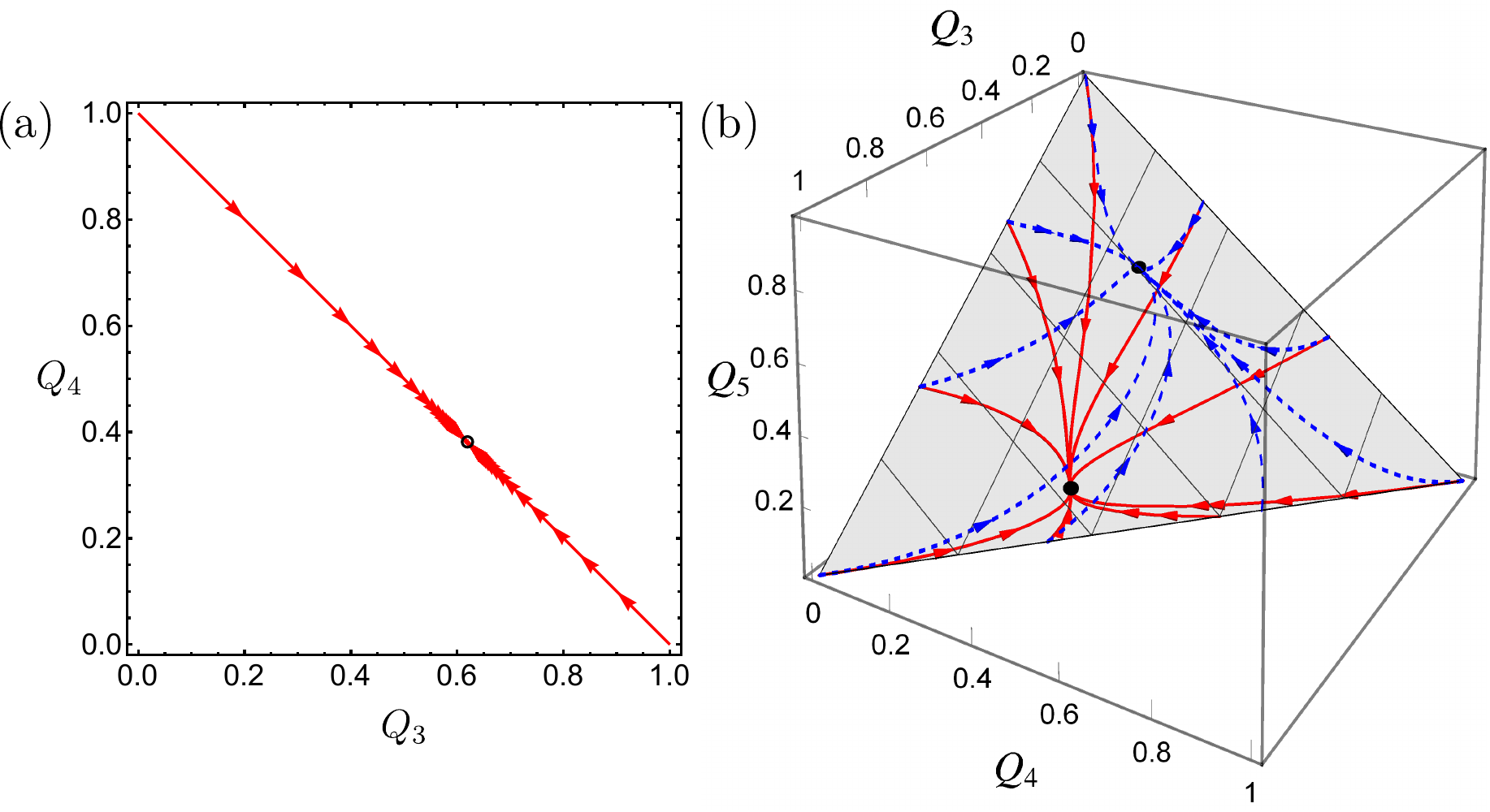}%
      \caption{
      The convergence to the unique steady state. (a) The dynamics of $3$-$4$ systems are confined to the line $Q_3 + Q_4 = 1$; here $p_{3,3}=q_{4,3}=1$. (b) The dynamics of $3$-$4$-$5$ system are confined to the plane $Q_3+Q_4+Q_5=1$. Illustrated are two sets of rates: $p_{3,3}=1$, $q_{4,3}=1$, $p_{3,4}=1$, $q_{5,3}=1$ (red) and $p_{3,3}=7$, $q_{4,3}=1$, $p_{3,4}=10$, $q_{5,3}=1$ (blue).
      }
      \label{fig:Approach}
   \end{figure}

\nin {\bf Exact solutions for $3$-$4$ systems:}
The simplest non-trivial systems to study, realisable under high-compression processes, consist only of $3$- and $4$-cells. Eqs. (\ref{eqmaster2}) then reduce to
\begin{eqnarray}
\dot{Q}_3 & = & (Q_3 - 2)\eta_{3,3} \nonumber \\
\dot{Q}_4 & = & (Q_4 + 1)\eta_{3,3} \nonumber \\
\dot{z} & = & -(z - 2)\eta_{3,3} \ .
\label{eq34evolution}
\end{eqnarray}
Since $Q_3+Q_4=1$ the first two equations are dependent. Integrating (\ref{eq34evolution}) and using (\ref{eqEta}) yields (see supplementary material)
\begin{equation}
t-t_0 = \frac{1}{p_{3,3}}\ln\left[\left(Q_3-a\right)^\alpha \left(Q_3-b\right)^\beta \left(Q_3-2\right)^\gamma \right] \ ,
\label{eqSoln34}
\end{equation}
with $a$ and $b$ the roots of $p_{3,3}Q_3^2 + q_{4,3}Q_3 - q_{4,3}$, and $\alpha=[(a-2)(a-b)]^{-1}$, $\beta=[(b-2)(b-a)]^{-1}$, and $\gamma=[(a-2)(b-2)]^{-1}$.
From this solution and (\ref{eq34evolution}) we can obtain $Q_4$ and $z$. 
Examples of these solutions are shown in Fig.~\ref{fig:example34Ini}. We observe that the steady state is the same regardless of the different initial states and is determined only by the rates.
As we show explicitly below, for $\mathcal{C}\leq 6$, this is not only a general feature but the steady state is also uniquely determined by the rates. This is illustrated for several $3$-$4$-$5$ systems in Fig.~\ref{fig:Approach}~(b) and discussed in more detail below.

\nin {\bf The steady state:}
The rightmost sum in (\ref{eqmaster2}) describes the rate of change of the total number of cells and it vanishes at steady state. 
A direct calculation of eqs. (\ref{eqmaster2}) for dense systems provides two significant observations: for systems consisting of highest cell order, $\mathcal{C}\leq 6$., (i) the steady state is unique; (ii) this steady state satisfies $\eta_{i,j}=0$ for all $i,j$, which implies a detailed balance-like state -- a very surprising result in systems that are manifestly far from conventional equilibrium.

For systems with $\mathcal{C}>6$, there are infinitely many steady state solutions, in addition to the detailed balance-like one (see details in the supplementary).
For example, steady states of systems with highest order $\mathcal{C}=7$ need only satisfy $\eta_{4,5}=\eta_{3,6}=0$ and $\eta_{4,4}=-\eta_{3,4}=-\eta_{3,5}=\eta_{3,3}$. Therefore, if $\eta_{3,3}\neq 0$, such steady states exist and they are not detailed balance-like.

\nin {\bf The approach to steady state:}
It is useful to understand the approach to the steady state, as well as its stability. Near this state, $Q_k(t)=Q_k^s + \delta Q_k(t)$ is only slightly different from its steady state value, $Q_k^s$, with $\lvert \delta Q_k \rvert \ll Q_k^s$. Defining $\vec Q(t) = \vec Q^s + \delta \vec{Q}(t)$, with $\vec Q \equiv (Q_3, \dots, Q_{\mathcal{C}})$, and expanding the r.h.s of eq. (\ref{eqmaster2}) to linear order, we obtain for $\delta \vec{Q}(t)$
\begin{align}
\delta \dot{\vec Q}(t) = A\cdot\vec Q(t) \ , \label{eq:linearisationGen}
\end{align}
in which the components of the constant matrix $A$ are cumbersome combinations of $p_{i,j}$, $q_{i,j}$, and the steady state fractions $Q_k^s$. 

Denoting the $i$th eigenvector and eigenvalue of $A$, respectively, by $\vec v_i$ and $\lambda_i$, we have near the steady state
\begin{align}
\vec Q(t) = \vec Q^s + \sum_i \, (\vec v_i^\dagger \cdot \delta \vec Q(t_0)) e^{\lambda_i t} \cdot \vec v_i%
\label{eq:tEvolSteadyState}%
\end{align}%
with some initial time $t_0$.
As we show in the supplementary material, all $\lambda_i$ must be real, there are no strictly positive eigenvalues and there has to be at least one negative eigenvalue.

To determine the unique steady-state solution of the ${3\text{-}4}$~system, we use the normalisation and detailed balance-like, $\eta_{3,3}^s = 0$, condition to find ${Q_3^s=[-\theta_{3,3}+\sqrt{\theta_{3,3}^2+4\theta_{3,3}}]/2}$, with $\theta_{i,j}\equiv q_{i+j-2,i}/p_{i,j}$. The normalisation confines the dynamics to a line in the $Q_3$-$Q_4$ plane, with the steady state as unique stable fixed point on it that is independent of the initial state. This was tested numerically for several systems and is illustrated in Fig. \ref{fig:Approach}(a), in which the arrow lengths represent the rate of approach to the steady state for $p_{3,3}=q_{4,3}=1$.

The dynamics of $3$-$4$-$5$ systems can be analysed similarly. Using $\eta_{3,3}^s=\eta_{3,4}^s=0$ and the normalisation condition, the steady state solution satisfies
   \begin{equation}
      (Q_3^s)^3 + \theta_{3,4}(Q_3^s)^2 + \theta_{3,3}\theta_{3,4}Q_3^s = \theta_{3,3}\theta_{3,4} \ , 
      \label{eq345Poly} 
   \end{equation}
and
   \begin{equation}
      Q_4^s = \frac{1}{\theta_{3,3}} Q_3^s \quad ; \quad Q_5^s = 1 - Q_3^s - Q_4^s \ .
   \end{equation}
Using the positivity of the $Q_k$, the uniqueness of the steady state can be established by studying the position of extrema of (\ref{eq345Poly}) \cite{WaThesis18_253} and it is determined by the rate fractions, $\theta_{i,j}$.
The normalisation confines the dynamics to the plane $Q_3+Q_4+Q_5=1$ and the approach to steady state in this plane is illustrated in Fig.~\ref{fig:Approach}(b) for two different sets of rates, which give two distinct steady states. 

For $\mathcal{C} \geq 6$, the dynamics take place on the hypersurface $\sum_k Q_k = 1$, on which there is at least one detailed balance-like steady state. As discussed above, for $\mathcal{C} > 6$, there may be additional steady states. Moreover, since all the steady states are stable, then, if they exist, they must be distinctly separate on the hypersurface.

\nin {\bf Simulations:} 
To test the theory, we carried out numerical simulations of quasi-static two-dimensional simple shear between parallel plates. 
We use a dissipative spheres model \cite{Bretal96,Sietal01} and the implementation of the model was provided by the open source project LIGGGHTS \cite{Kletal12}.
Within this implementation, we used a Hertz force model for the grain-grain and grain-walls interactions, with: Young's Modulus $5\times10^{6}$Pa, restitution coefficient $0.2$, Poisson's ratio $0.5$, and friction coefficient $0.5$.
The system consisted of $N=21,690$ spheres restricted to move in the plane $z=0$ of four different diameters, $7$mm ($5402$ grains), $9$mm ($5400$), $11$mm ($6120$) and $14$mm ($4770$), respectively.

The system has a length of $1.65$m and periodic boundary conditions in the $x$-direction. The initial state was generated by compressing the grains from an initially loose random distribution by a flat surface at constant pressure, $P_\mathrm{initial}=54.45$Kg/m, in the $y$-direction until their total kinetic energy fell below $10^{-12}$J per particle. This state was regarded as mechanically stable. Then, maintaining the confining pressure, shear started in the $x$ direction, with shear velocity $v_{\gamma}=0.06$m/s. No gravitational field was applied.

The time-step was $10^{-6}$s and the grain positions and velocities were saved every $200$ time-steps for collecting detailed data on the contact network evolution.\\
At each stop, cells and the grains surrounding them were identified, from which we tracked the evolution of the COD and obtained the rates of contact events, $p_{ij}$ and $q_{ij}$. Cells of orders higher than $13$ occurred very rarely and, therefore, were excluded from the analysis.

\begin{figure}[htbp]
   \centering
      \includegraphics[width=.48\textwidth]{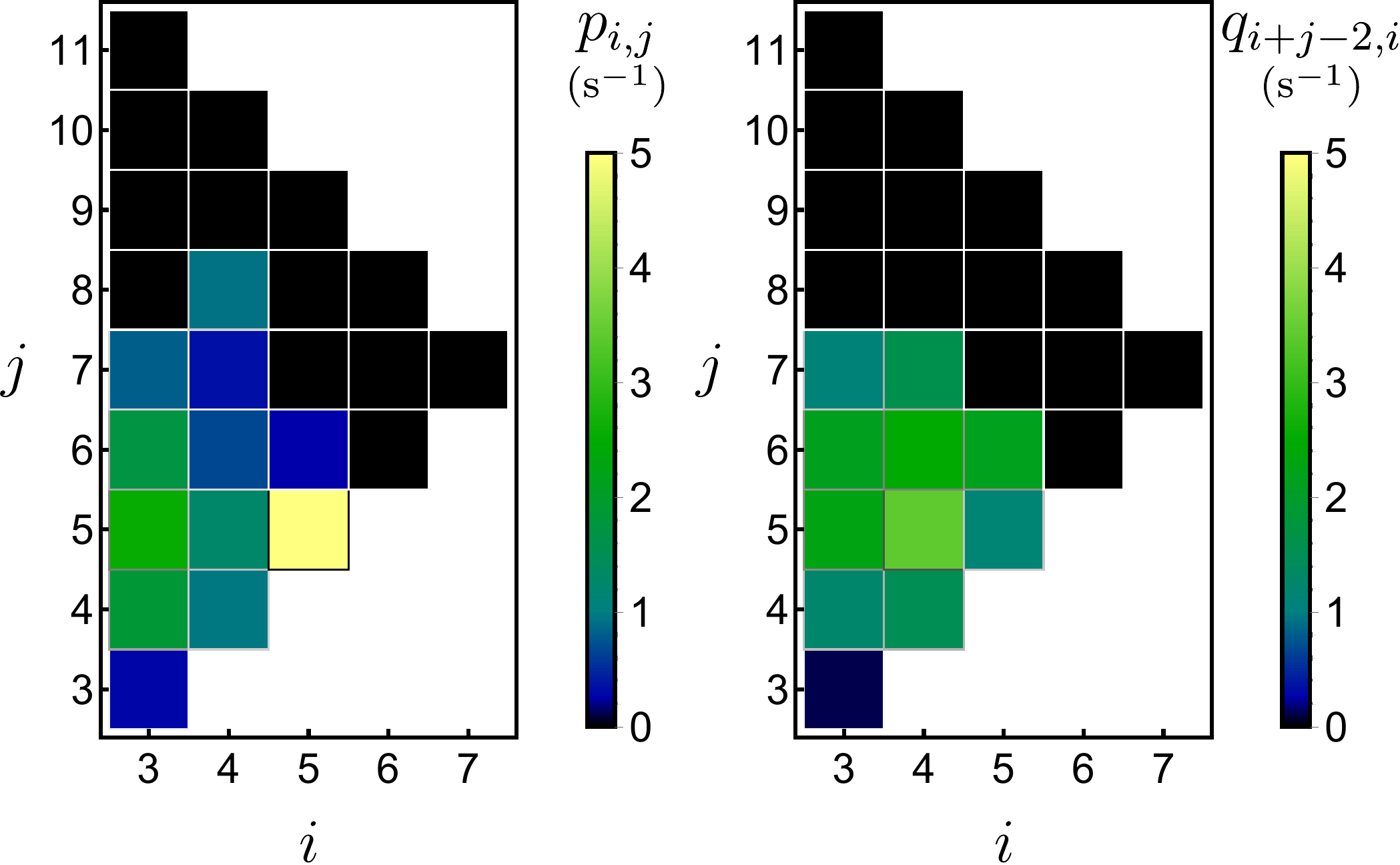}%
   \caption{The rates $p$ and $q$, computed from the simulation, are used to solve the equations. Note that $p_{i,j}$ and $q_{i+j-2,i}$ are symmetric under exchanging $i$ and $j$.}
   \label{fig:rates}
\end{figure}

\nin {\bf Analysis of simulation data:}
The rates are determined from the numbers, $N^{(p)}_{i,j}\equiv p_{i,j} \, Q_i Q_j N_c \Delta t$ and $N^{(q)}_{i,j}\equiv q_{i+j-2,i}\, Q_{i+j-2} N_c \Delta t$, of the events ${i+j\xleftrightharpoons{} i+j-2}$ during a time interval $\Delta t$.
In general, the rates should depend on the force distribution and the shear rate, both of which change slightly during the simulation as the system dilates before reaching a steady state. Since the above analysis is for constant rates, this could complicate a direct comparison between the analytical solutions and the simulation data. Fortunately, the rates change little and their time dependence can be neglected.
Fewer than $15$ cells of a specific CO or fewer than $40$ events per $0.1$s of a specific process $\eta_i,j$ lead to large statistical errors and were ignored.
Interestingly, in spite of the long simulation time, the rates fluctuated significantly, $30\%-65\%$, perhaps because of the sensitively to small changes in $Q_k$. Smaller fluctuations may be achieved in larger systems with more contact events.
Fig. \ref{fig:rates} shows the rate diagram, calculated from data collected during intervals of $0.1$s and averaged over the entire evolution process. Such rate diagrams characterise uniquely a dynamic process.

\begin{figure}[htbp]
   \centering
   \includegraphics[width=.4\textwidth]{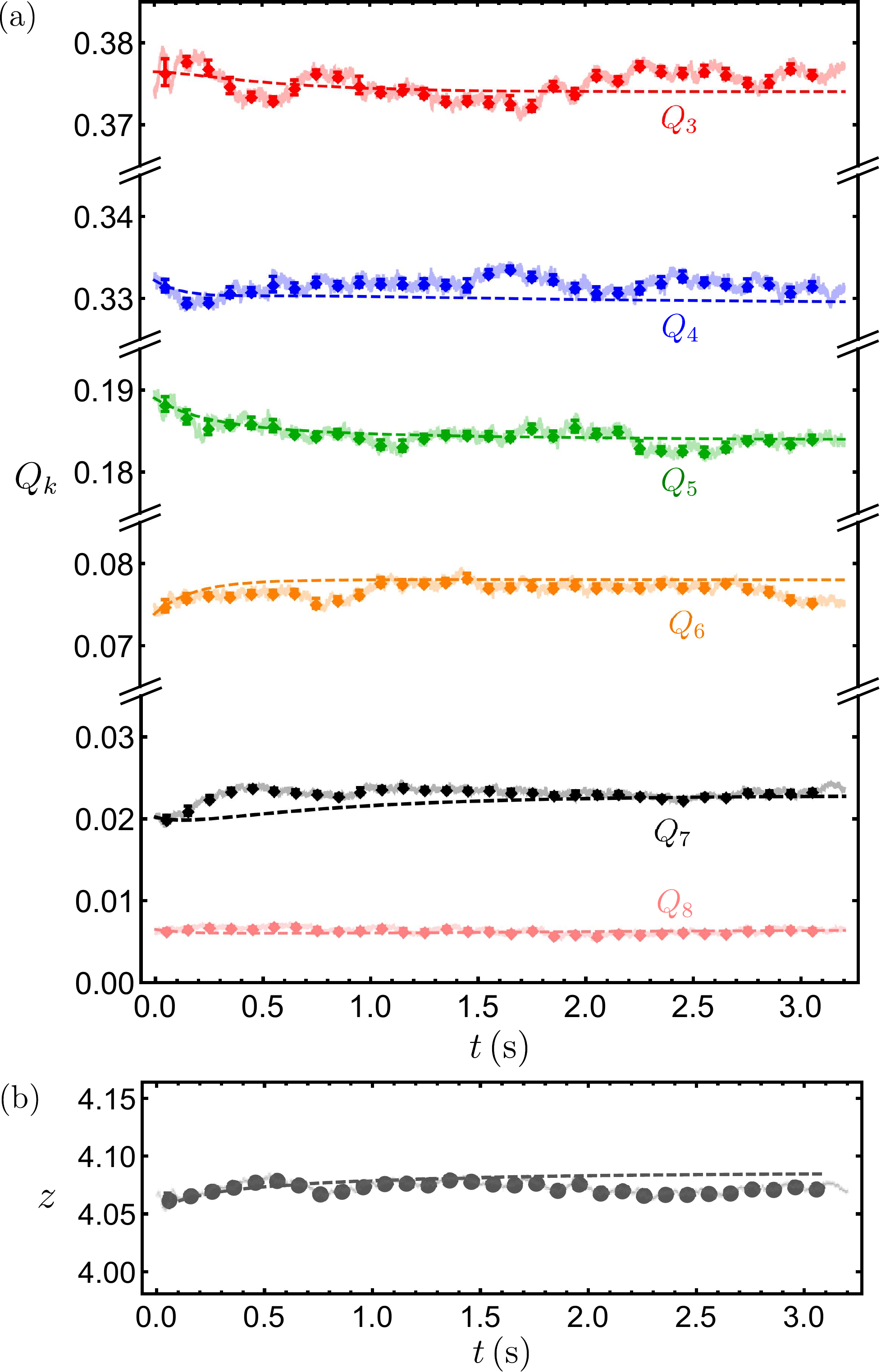}
   \caption{
   The evolution of (a) the COD for cell orders $3$-$8$ (from top to bottom) and (b) the mean coordination number, $\bar{z}=2e/(e-2)$ with $e=\sum_k k Q_k$. The continuous simulation results (light lines) agree with the shown averages taken over intervals of $0.1$s. The error bars represent the standard deviation during those intervals. The solution of equations (\ref{eqmaster2}) (dashed lines) agrees well with the simulation results.}
   \label{fig:numericalSolution}
\end{figure}

Using the calculated rates, shown in Fig. \ref{fig:rates}, we solved the evolution equations (\ref{eqmaster2}) numerically, with the initial COD determined by averaging over the first $0.1$s in the simulation. While there was a reasonably good agreement between the solution and the simulation data when using the computed mean rates, we found that an even better agreement was achieved by correcting only $q_{6,4}$ by $10\%$. Note that this correction is much smaller than the aforementioned statistical fluctuations of the rates. 
The good agreement between our solution and the simulation data after this minor correction is shown in Fig. \ref{fig:numericalSolution}.

\nin {\bf Conclusion:}
To conclude, we have constructed master equations to describe the evolution of a key structural characteristic of granular media - the cell order distribution (COD) - from any initial state, given the cell merging and splitting rates. The equations yield a surprising result: the steady states of the non-equilibrium dynamics of granular systems, with cell order no higher than $\mathcal{C}=6$ satisfy a detailed balance-like condition. This detailed balance-like steady state is unique and stable. 
Systems including cell orders of $7$ and higher can converge to this steady state, as well as to other solutions that do not support detailed balance, all of which are {\it stable} fixed points of the equations.

We validated the theory by running a long simulation of a sheared system, determining the rates, and using those to solve the master equations. Indeed, the calculated solution agrees nicely with the simulated COD evolution, in spite of large fluctuations in the rates computed from the simulations. These fluctuations may be caused by `clappers': Particle pairs that make and break their common contact repeatedly. We plan an experiment to study the relative contributions of clappers and non-clappers to the rate statistics.

In our analysis, we assumed constant rates throughout the dynamic process. This is unlikely to be the general case and extending the theory to time-dependent rates is the next step. Such time dependence is expected because the contact event rates is likely to be sensitive to the intergranular force distribution, which is position- and time-dependent. 
Indeed, it has been argued \cite{Bletal15,MaBl17} that the structural evolution of granular system is a self-organisation process coupled to the evolution of the intergranular forces. Therefore, these equations and their extension form an important step towards a complete understanding of the structure-forces co-evolution. 
Another extension would be to open systems, when the number of particles is not conserved. This extension could be related to a grand-canonical statistical mechanical description of granular ensembles \cite{Bl16}. 
Although the evolution eqs. (\ref{eqmaster1}) include effects of rattlers, we disregarded these in the analysis for clarity. Indeed, very dense systems with low values of $\mathcal{C}$ should be almost entirely rattler-free. It should be emphasized that including gravity and body forces obviates the rattler terms in eqs. (\ref{eqmaster1}) because then all the grains transmit forces and make cells, with the rattlers making mainly low orders ones.
Finally, the theory can be extended to three-dimensions, in which case the main hurdle - cell classification -  can be overcome using existing methods \cite{Kraynik,Jordan}.

\nin {\bf Acknowledgements:}
C.C.W. acknowledges the Erasmus+ programme of the EU and the Studienstiftung des Deutschen Volkes (German Academic Scholarship Foundation) for funding the visit to the Cavendish Laboratory, where this work was done.

   \appendix
   
   \clearpage

   \begin{center}
   \widetext
   \large\bf Structural Evolution of Granular Systems: Theory\\[2mm]Supplementary Materials \\[\baselineskip]
   \end{center}
   \twocolumngrid
   
   \setcounter{equation}{0}
   \setcounter{figure}{0}
   \setcounter{table}{0}
   \setcounter{page}{1}
   
\nin{\bf The approach to steady state:}
In the section about the approach to steady state we claimed that ${\sum_k \delta Q_k = 0}$. To establish this identity, we start from the full linear expansion of the evolution equations sufficiently close to steady state:
   \begin{widetext}
   \begin{alignat}{2}
      \delta \dot Q_k
         & =
             && \frac{1}{2} \,
                \sum_{i=3}^{k-1} \left[
                   p_{i,k-i+2} \left(Q_{k-i+2}^s\delta Q_i+Q_i^s\delta Q_{k-i+2}\right) - q_{k,i} \delta Q_k
                \right] (1+\delta_{i,k-i+2}) \notag \\
         & && - \sum_{i=k+1}^{i_\mathrm{m}} \left[
                   p_{k,i-k+2} (Q_k^s\delta Q_k+Q_{i-k+2}^s \delta Q_{i+k-2}) - q_{i,k} \delta Q_i
                \right] (1+\delta_{i,k-i+2}) \notag \\
          & - Q_k^s\sum_{\ell=3}^\mathcal{C}
             &&
                 \left\{\vphantom{\sum_{i=\ell+1}^{i_\mathrm{m}}}\right.
                 \frac{1}{2} \,\sum_{i=3}^{\ell-1} \left[
                   p_{i,\ell-i+2} \left(Q_{\ell-i+2}^s\delta Q_i+Q_i^s\delta Q_{\ell-i+2}\right) - q_{\ell,i} \delta Q_\ell
                \right] (1+\delta_{i,\ell-i+2}) \notag \\
         & && - \sum_{i=\ell+1}^{i_\mathrm{m}} \left[
                   p_{\ell,i-\ell+2} (Q_\ell^s\delta Q_\ell+Q_{i-\ell+2}^s \delta Q_{i+\ell-2}) - q_{i,\ell} \delta Q_i
                \right] (1+\delta_{i,\ell-i+2})
                \left.\vphantom{\sum_{i=\ell+1}^{i_\mathrm{m}}}\right\}\notag \ ,
   \end{alignat}
   with $i_m\equiv\left\lvert\left(\frac{\mathcal{C}+2}{2}\right)\right\rvert$. Summing over the index $k$ and using $\sum_k Q_k=1$, we obtain
   \begin{alignat}{2}
      \sum_{k=3}^\mathcal{C}\delta \dot Q_k
         & = \hphantom{\sum_k Q_k}\sum_{k=3}^\mathcal{C}
             && \left\{\vphantom{\sum_k^C}\right.\frac{1}{2} \,
             \sum_{i=3}^{k-1} \left[
                   p_{i,k-i+2} \left(Q_{k-i+2}^s\delta Q_i+Q_i^s\delta Q_{k-i+2}\right) - q_{k,i} \delta Q_k
                \right] (1+\delta_{i,k-i+2}) \notag \\
         & && - \sum_{i=k+1}^{i_\mathrm{m}} \left[
                   p_{k,i-k+2} (Q_k^s\delta Q_k+Q_{i-k+2}^s \delta Q_{i+k-2}) - q_{i,k} \delta Q_i
                \right] (1+\delta_{i,k-i+2}) \notag
                \left.\vphantom{\sum_k^C}\right\}
                \\
          & - \underbrace{\sum_k Q_k^s}_{=1}
             \sum_{\ell=3}^\mathcal{C}
             &&
                 \left\{\vphantom{\sum_{i=\ell+1}^{i_\mathrm{m}}}\right.
                 \frac{1}{2} \,\sum_{i=3}^{\ell-1} \left[
                   p_{i,\ell-i+2} \left(Q_{\ell-i+2}^s\delta Q_i+Q_i^s\delta Q_{\ell-i+2}\right) - q_{\ell,i} \delta Q_\ell
                \right] (1+\delta_{i,\ell-i+2}) \notag \\
         & && - \sum_{i=\ell+1}^{i_\mathrm{m}} \left[
                   p_{\ell,i-\ell+2} (Q_\ell^s\delta Q_\ell+Q_{i-\ell+2}^s \delta Q_{i+\ell-2}) - q_{i,\ell} \delta Q_i
                \right] (1+\delta_{i,\ell-i+2})
                \left.\vphantom{\sum_{i=\ell+1}^{i_\mathrm{m}}}\right\}\notag \ .
   \end{alignat}
   Renaming the index $\ell$ as $k$, the individual sums neatly cancel out:
   \begin{alignat}{2}
         \sum_{k=3}^\mathcal{C} \delta \dot Q_k
         & = \sum_{k=3}^\mathcal{C}
             && \left\{\vphantom{\sum_k^C}\right.\frac{1}{2} \,
             \sum_{i=3}^{k-1} \left[
                   p_{i,k-i+2} \left(Q_{k-i+2}^s\delta Q_i+Q_i^s\delta Q_{k-i+2}\right) - q_{k,i} \delta Q_k
                \right] (1+\delta_{i,k-i+2}) \notag \\
         & && - \sum_{i=k+1}^{i_\mathrm{m}} \left[
                   p_{k,i-k+2} (Q_k^s\delta Q_k+Q_{i-k+2}^s \delta Q_{i+k-2}) - q_{i,k} \delta Q_i
                \right] (1+\delta_{i,k-i+2})\left.\vphantom{\sum_k^C}\right\} \notag \\
          & - \sum_{k=3}^\mathcal{C}
             && \left\{\vphantom{\sum_k^C}\right.\frac{1}{2} \,
             \sum_{i=3}^{k-1} \left[
                   p_{i,k-i+2} \left(Q_{k-i+2}^s\delta Q_i+Q_i^s\delta Q_{k-i+2}\right) - q_{k,i} \delta Q_k
                \right] (1+\delta_{i,k-i+2}) \notag \\
         & && - \sum_{i=k+1}^{i_\mathrm{m}} \left[
                   p_{k,i-k+2} (Q_k^s\delta Q_k+Q_{i-k+2}^s \delta Q_{i+k-2}) - q_{i,k} \delta Q_i
                \right] (1+\delta_{i,k-i+2})\left.\vphantom{\sum_k^C}\right\} \notag\\
         & = 0
                \text{.} && \label{eq:linMasterSummed}
      \end{alignat}
   \end{widetext}

Furthermore, the cell fractions, $Q_k(t_0)$, must be normalised at all times and, in particular, close to steady state. It follows that $\sum_k \delta Q_k(t_0)=0$. Since $\sum_k \dot \delta \dot Q_k = 0$, as the above calculation shows, we obtain that $\sum_k \delta Q_k(t) = 0$ for all times $t$ in the linearised regime as well. \\
   
\begin{figure}[htbp]
      \centering
      \subfigure[]{
         \includegraphics[width=.45\textwidth]{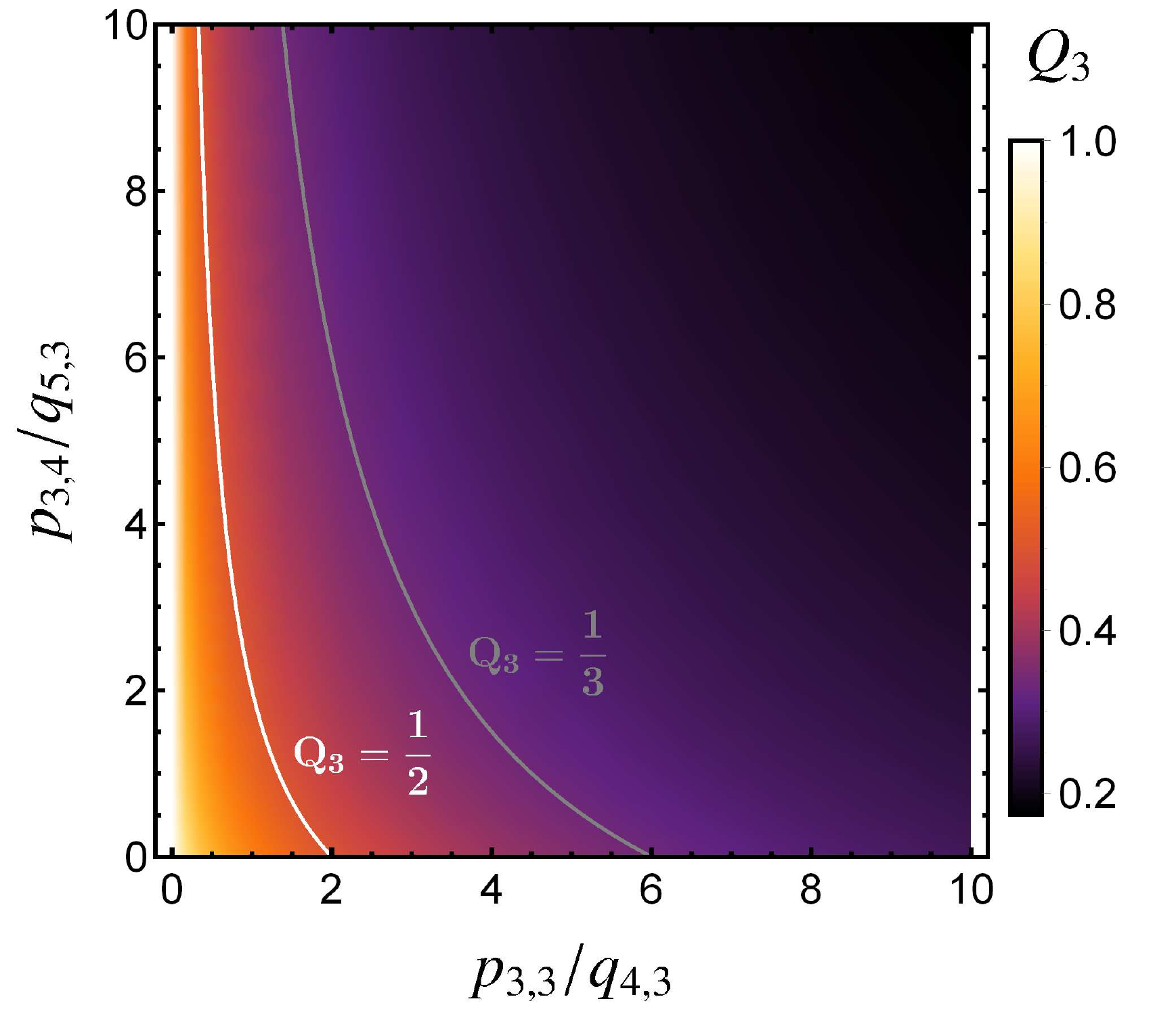}
         \label{fig:phaseDiagA}
      }
      \subfigure[]{
         \includegraphics[width=.45\textwidth]{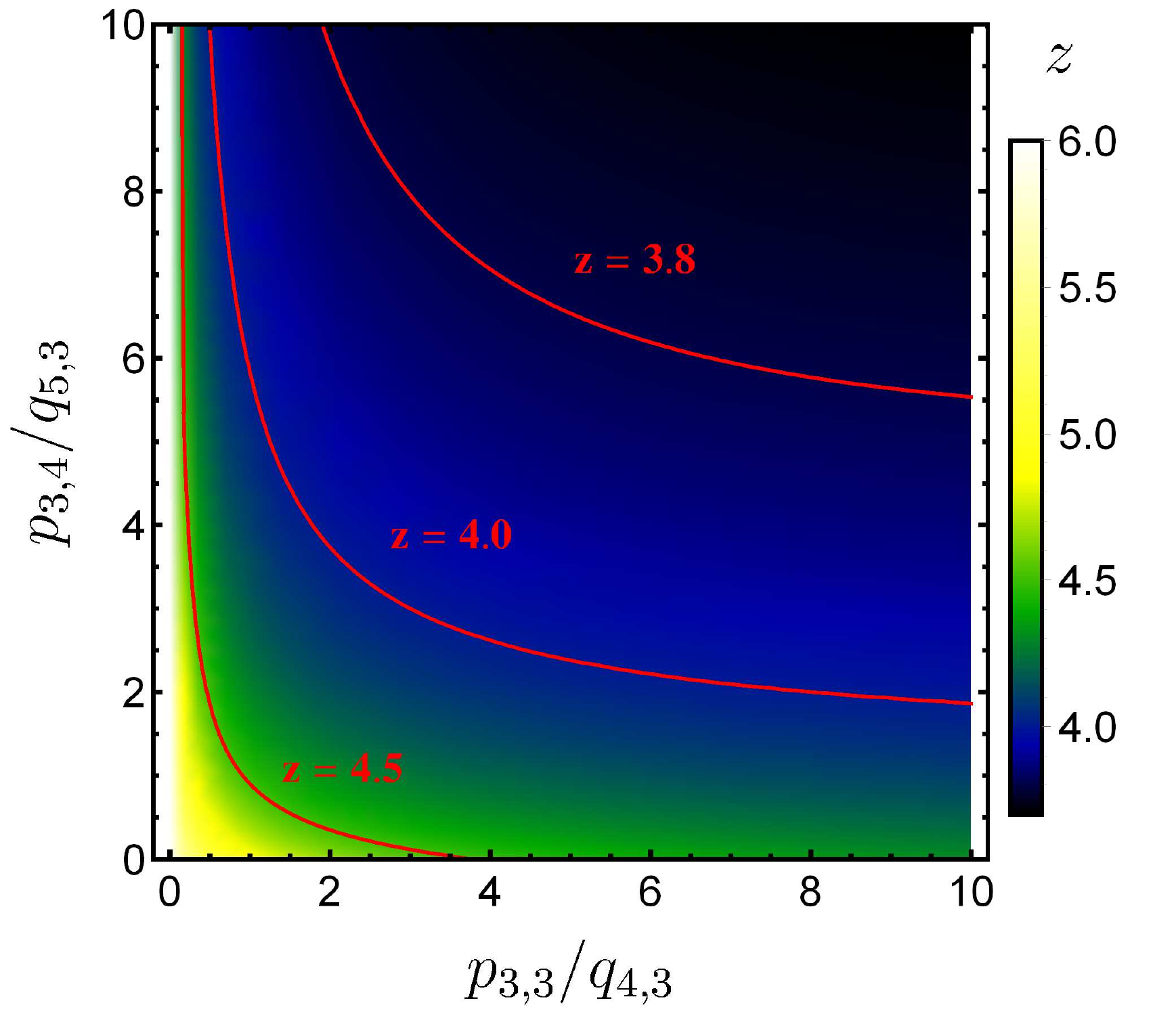}
         \label{fig:phaseDiagB}
      }
      \caption{Phase diagrams of the steady-state values of (a) $Q_3$ and (b) $z$, for the $3$-$4$-$5$ system, plotted in the phase space spanned by $p_{3,3}/q_{4,3}$ and $p_{3,4}/q_{5,3}$.}
      \label{fig:phaseDiag}
   \end{figure}

\nin{\bf Phase diagram:}
For $\mathcal{C}\leq 6$ the steady state can be uniquely expressed in terms of the fractions of the occurring rates for reaction and back-reaction, $p_{i,j}/q_{i+j-2,i}$. Therefore, multiplying all the rates by a constant factor does not affect the steady state, but only modifies the time by which the steady state is reached. This allows us to scale the rates and represent the steady state fractions, $Q_k$ and the mean coordination number, $z$, as contours in a phase diagram in a phase space spanned by the rate fractions. 
In particular, the phase diagram of a system, containing only cells of orders $3$, $4$ and $5$, can be conveniently represented by a density plot, see Fig.~\ref{fig:phaseDiag}. Fig. \ref{fig:phaseDiag} shows phase diagrams of (a) the ultimate fraction of $3$-cells and (b) the mean coordination number $z$, in the phase space spanned by the rates. Such phase diagrams demonstrate how the master equations can provide guidelines to design specific packing protocols.
   
\nin{\bf Clapping events:}
A particular phenomenon, observed extensively in all the simulations we ran, is the occurrence of `clapping' events. These are contact events involving pairs of particles making and breaking the contact between them repeatedly. 
For simplicity, we defined non-clapping events in our analysis as events that occur only once during the system's dynamics. However, since simulations and experiments take place over finite duration, non-clapping events are defined practically as events that occur only once during the simulation/experiment. 
This provided us with a straightforward way to identify and distinguish between the two types of events in our simulations. 

Evidently, where clapping occurs, the grains did not move away from one another, which may mean that the local configuration has not changed much. This implies that clapping events only add noise to the statistics of events and evolution of the COD.
The probability that grains have moved away from one another and have come back together at a later time is expected to be very low and we have not observed any such event for the 500 clapping pairs, which we tracked explicitly during the entire simulation. \\
A potential problem with the above definition is that events occurring close to the end of the simulation could be identified wrongly as non-clapping because the grains do not have the time to clap. This should not affect the statistics much because it means that half a clap has been counted as a non-clapping event and we just missed the other half of the clap. This would be significant only if the clap affects two cell types whose splitting or merging is rare. However, to avoid this potential error, reliable results should be obtained sufficiently early and far away from the end of the simulation.

It is important to note that clapping does not contribute to the structural evolution on a coarse grained time scale, which the master equations come to describe. Clapping obeys `detailed balance' in the sense that the number of `reactions' equals the number of `back-reactions' and $\eta_{i,j}^\mathrm{clap}=0$. 
Furthermore, the master equations, $\dot Q_k = f(\{\eta_{i,j}\})$, are linear in the $\eta$s, $\dot Q_k = f(\{\eta_{i,j}^\mathrm{non-clap}\})+f(\{\eta_{i,j}^\mathrm{clap}\})$. This means that the contribution of the clapping events to $\dot{Q}$ in the master equations vanishes on a coarse-grained time scale that is sufficiently longer than the time between clapping events.
We found in our simulations that most clapping pairs were active for some duration, went dormant, and became active again (see Fig. \ref{fig:clappingPairs} below). Whether this is a result of the specific dynamics (e.g. shear, compression, etc.) is a question we intend to explore in future work. \\

\nin{\bf Discrete Element Method (DEM) simulation:} %
The main text describes simple shear simulations. To study another dynamic process, we analysed a long-time DEM simulation of a bi-axial compression.
The simulated system consisted of $22,381$ discs, whose diameters were distributed log-normally \cite{MaBl14,MaBl17}. The discs interaction was harmonic, with tangential and normal spring constants $k_\mathrm{t}$ and $k_\mathrm{n}$, respectively, obeying $k_\mathrm{t}/k_\mathrm{n}=1/4$. The mean overlap at the contacts is estimated as $d/D=\sigma_c/k_n=10^{-5}$, where $\sigma_c$ is the initial confining pressure. This overlap is much smaller than the one used for the simple shear simulation. The restitution coefficient was set to $0.98$ -- much larger than in the simple shear simulations. The initial configuration was prepared with a very low inter-particles friction. At the start of the simulation, the particles were assigned a friction coefficient of $\mu=0.5$, the same as in the simple shear simulation.
The system was periodic in both dimensions and was subjected to a gradual vertical compression keeping the lateral confinement pressure constant. This resulted in a quasi-static dilation with a diagonal shear.

\begin{figure}[h]
   \centering
   \includegraphics[width=.48\textwidth]{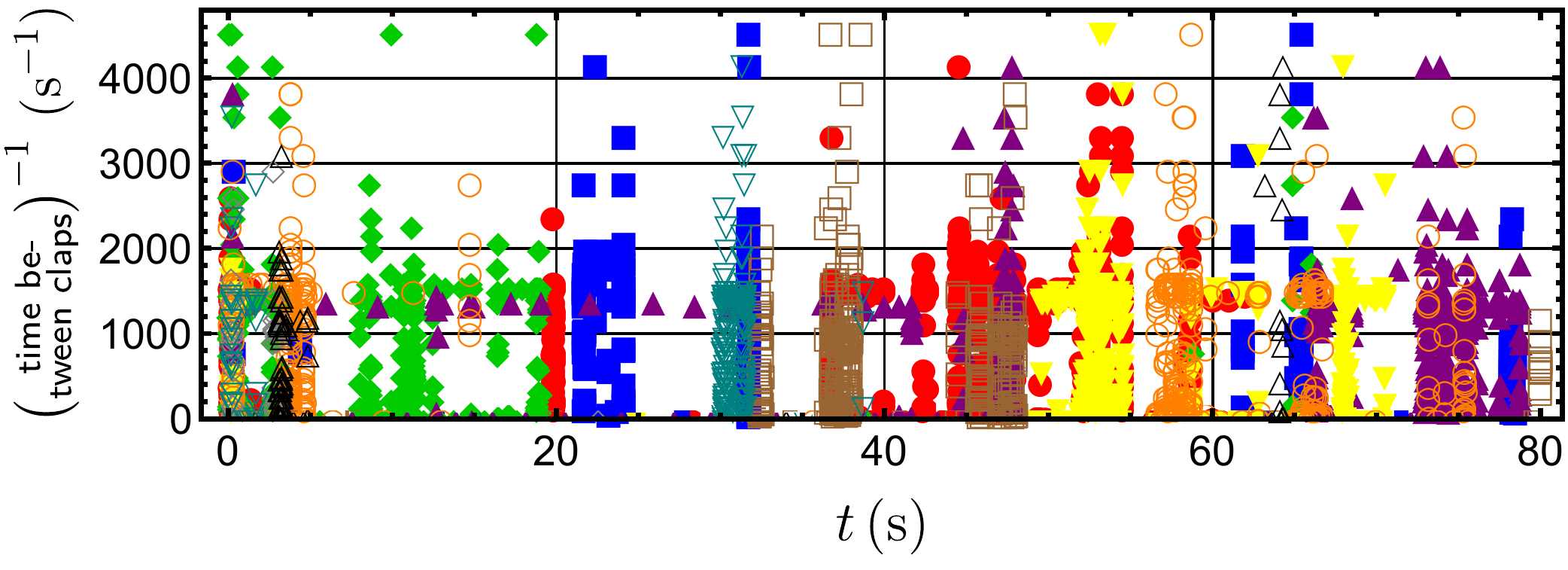}
   \caption{$10$ clapping pairs traced throughout the time evolution: each colour indicates a specific pair of grains. The $y$-axis shows the inverse time between subsequent claps. It can be observed that some clapping pairs, e.g. the blue and orange pairs, are active for a while, lay dormant and become active again at a later stage.}
   \label{fig:clappingPairs}
\end{figure}

The particle stiffness and the high restitution coefficient enhance clapping, making it a good system to study this phenomenon. We chose to trace $448$ individual clapping pairs throughout the simulation, as tracing all pairs would have been too time-consuming. 
We found that a pair can indulge in repeated clapping, disengage for a while, and become active again later. This can be seen in Fig. \ref{fig:clappingPairs}, in which we show the inverse time between claps as function of the time in the simulation for $10$ arbitrarily picked pairs. Clapping dominated the contact events in this simulation: only $0.7\,\%$ of all events were non-clapping, suggesting that the structure is hardly changed within large regions. In this respect, a dilation process is quite different from simple shearing in that it does not mix the system quite as well.

\begin{figure}
   \centering
   \includegraphics[width=.48\textwidth]{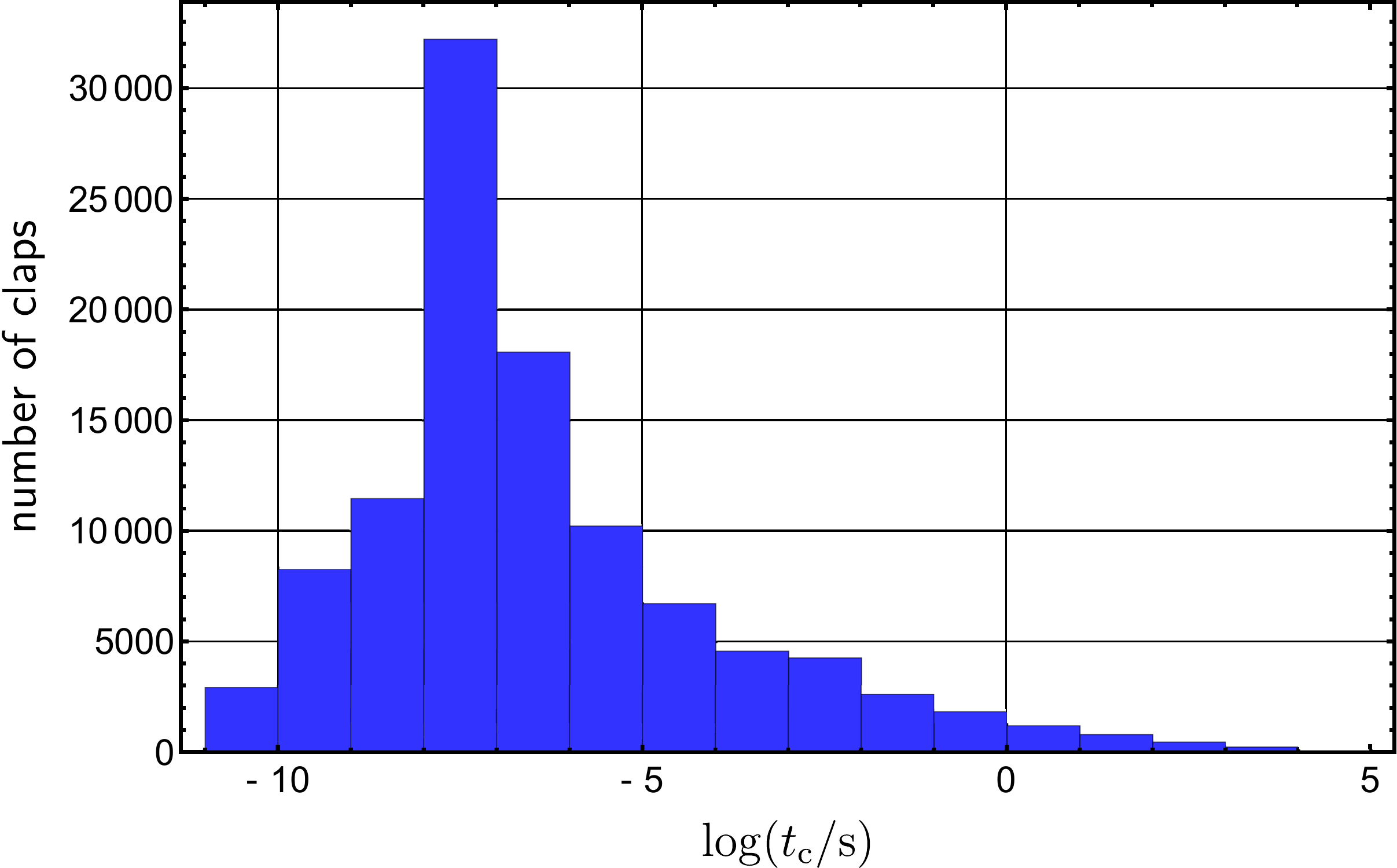}
   \caption{Distribution of the time between claps $t_c$: short times can be associated with claps in one sequence while large times correspond to the separation between subsequent cascades of claps.}
   \label{fig:distClapTimes}
\end{figure}

\begin{figure}[htbp]
   \includegraphics[width=.49\textwidth]{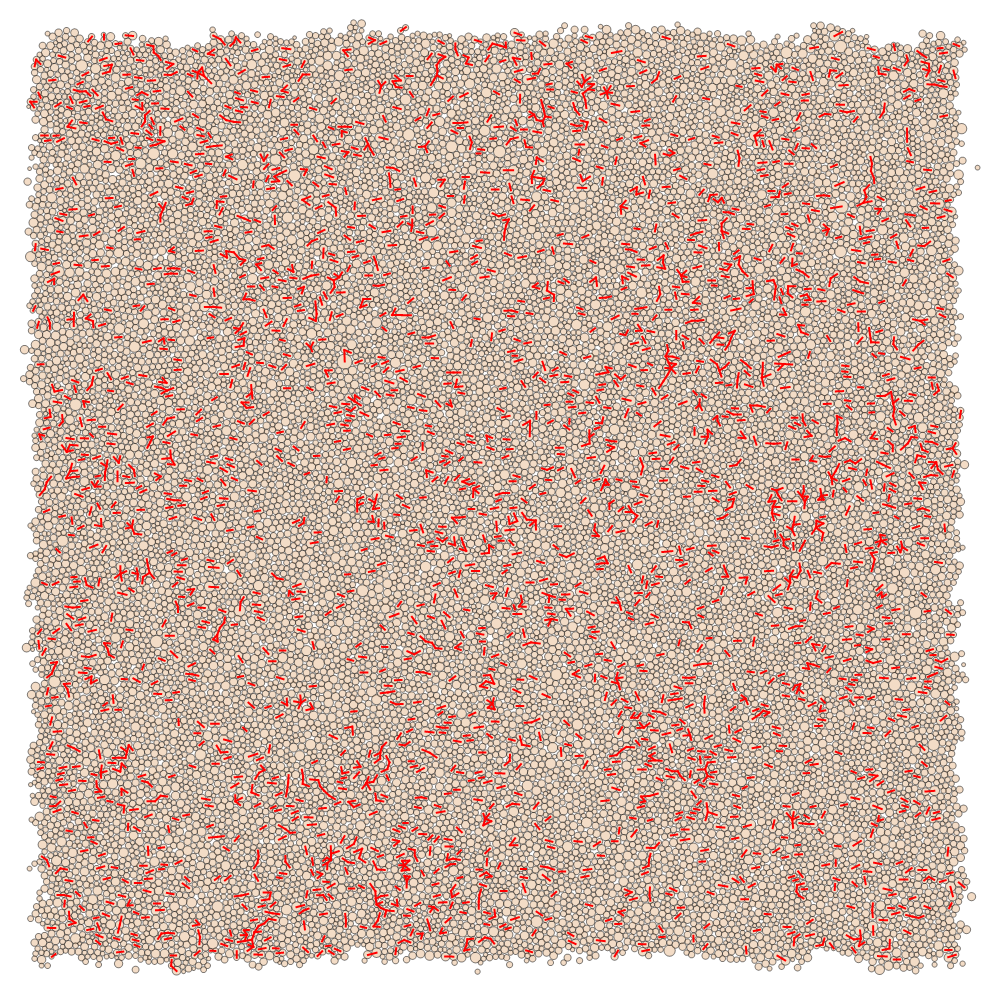}
   \caption{Location of non-clapping events in the system: events taking place between $t=0.04\,\mathrm{s}$ and $t=0.06\,\mathrm{s}$ are signified by a red line connecting particle centres of the grains involved in the CE.}
   \label{fig:locationNonClap}
\end{figure}

The wide distribution of the time intervals between claps of a specific pair is evident from the very long tail of the histogram in Fig. \ref{fig:distClapTimes}. 
The clapping seem also to be spatially uncorrelated, occuring uniformly across the system, as shown in Fig. \ref{fig:locationNonClap}.

Unfortunately, the dilation simulation does not reach a steady state in which the rates remain constant. The most likely reason is that the boundary forces change continually with time, which changes the statistics of the intergranular contact force distribution, in turn affecting the rates. Indeed, the rates appear to decay with time exponentially, $\sim e^{-t/\tau}$, with $\tau$ ranging from $0.2$ to $2.2$, depending on the specific rates. Since the theory presented is limited to constant rates, we have not pursued a detailed modelling of the COD evolution of this simulation. The continual decay of rates, in spite of the long time simulation, further suggests that a steady state may not be reached before the dilation disrupts mechanical equilibrium. This highlights the fact that not every process reaches necessarily a steady state. \\

\end{document}